\newcommand{\magcir}{\raise -2.truept\hbox{\rlap{\hbox{$\sim$}}\raise5.truept
\hbox{$>$}\ }}

\documentclass{PoS}

\title{The AGN component in deep radio fields: \\ 
Results from the First Look Survey}

\ShortTitle{The AGN component in the First Look Survey}

\author{\speaker{Isabella Prandoni}\thanks{This work has been partly 
performed at ASTRON under the Helena Kluyver female visitor programme}\\

        INAF - Istituto di Radioastronomia, Via P. Gobetti 101, 40129 Bologna, 
Italy\\
        E-mail: \email{prandoni@ira.inaf.it}}

\author{Raffaella Morganti\\

        Netherlands Institute for Radio Astronomy, Postbus 2, 7990 AA, 
Dwingeloo, NL \\
Kapteyn Astronomical Institute, University of Groningen, Postbus 800, 9700 AV 
Groningen, NL }

\author{Arturo Mignano\\
        INAF - Istituto di Radioastronomia, Via P. Gobetti 101, 40129 Bologna, 
Italy}

\abstract{We are currently exploiting the deep radio/optical/IR information 
available for the extra-galactic component of the Spitzer First Look Survey 
(FLS) to investigate the physical properties of faint radio-selected AGNs, 
with
the aim of studying the AGN component of sub--mJy radio fields. 
One of the key unresolved issues is whether, as a function of cosmic epoch, 
low-power AGNs are more related to efficiently accreting systems (mostly 
radio-quiet) or to systems with very low accretion rates (mostly radio-loud). 
Here we present a sample of optically 
identified radio-emitting AGNs extracted from the FLS.
Preliminary results show that at the flux densities probed by the FLS 
($S_{\rm 1.4 GHz}\magcir 100$ $\mu$Jy) we still have a significant number of 
radio--loud AGNs, similarly to what found in 'brighter' sub--mJy radio 
samples. Very interestingly, however, we have also a clear and direct 
evidence of a population of radio--emitting 
AGNs in the FLS, whose properties are consistent with those 
expected from existing radio--quiet AGN modeling. Such AGNs could be 
recognised as such thanks to the availability 
of IR colors which proved to be especially useful to 
efficiently separate radio sources triggered by AGNs, from sources 
triggered by star-formation. 
This latter result supports the idea
that radio--quiet AGNs are not necessarily radio silent, and very promisingly
may indicate that the bulk of the radio--quiet 
AGN population could emerge from studies of deeper radio samples.}

\FullConference{Panoramic Radio Astronomy: Wide-field 1-2 GHz research on 
galaxy evolution - PRA2009\\
		 June 02 - 05 2009\\
		 Groningen, the Netherlands}

\begin{document}

\section{Scientific background}
After a decade of multi-wavelength studies of deep radio fields 
it is rather clear that star-forming galaxies dominate at 
microJy ($\mu$Jy) levels, while radio sources associated to 
early--type galaxies, and plausibly triggered by AGNs, are the most
significant source component at flux densities $>50-100$ $\mu$Jy 
($\sim 60-70\%$ of the total) with a further 10\% contribution from
broad-/narrow-line AGNs 
(see e.g. \cite{Gru99,Pra01,Afo06,Mig08,Smo08}). \\
The somehow unexpected presence of large numbers of AGN--related sources at 
sub-mJy fluxes has given a new interesting scientific perspective to the 
study of deep radio fields. Of particular interest is the possibility of 
studying the physical and evolutionary properties of such low power/high 
redshift AGNs. One of the key unresolved issues 
is for instance whether, as a function of cosmic epoch, the low-power AGNs 
are more related to efficiently accreting systems - like 
radio-intermediate/quiet quasars - or to systems with very low accretion rates 
- like e.g. FRI radio galaxies (\cite{Fan74}). The 
latter scenario (radio mode) is supported by the presence of many optically 
inactive early type galaxies among the sub--mJy radio sources; whereas the 
quasar mode scenario may be supported by the large number of so-called 
radio-intermediate quasars observed at mJy levels (see e.g. \cite{Lac01})
and by the modeling work of \cite{Jar04}, who predict large numbers
od radio-quiet AGNs at sub-mJy levels.\\
Assessing such a question would have relevant impact on topics 
like: the role played by low accretion/radiative efficiency AGNs in the global 
black-hole-accretion history of the Universe; the relative contribution of 
radiative versus kinetic (jet-driven) feedback to the global AGN feedback in 
models of galaxy formation; and, more generally, would allow us a better 
understanding of the triggering mechanisms of AGN radio activity. 

\section{The First Look Survey}
The First Look Survey was the first major scientific program carried 
out by the Spitzer Space Telescope. 
As part of the extragalactic component of the First Look Survey (FLS), 
a region covering 4 square degrees and centered on RA=17:18:00, 
DEC=59:30:00 was imaged, with the aim of studying a low Galactic Background 
region to a significantly deeper level than any previous large-area 
extragalactic infrared survey. 
Such a survey was complemented by a smaller ~0.75x0.3 sq. degr. 
survey ('verification' survey), lying in the same region, observed 
to a factor of 3 deeper flux levels. \\
Spitzer images and source catalogues are available at 3.6, 4.5, 5.8, 8.0 
$\mu$m (IRAC, \cite{Lac05}) and at 24, 70, 160 $\mu$m (MIPS, 
\cite{Fad06,Fra06}), complemented by a large set of ancillary data taken 
at different wave-bands: from deep optical imaging (R-band, \cite{Fad04}; 
u*-/g*-bands, \cite{Shi06}) and spectroscopy (\cite{Sto02,Pap06,Mar07})
to deep (rms noise level~23-30 muJy)
radio images at 1.4 GHz (VLA, \cite{Con03}) and 610 MHz (GMRT, \cite{Gar07}). 
A deeper (rms noise level ~8.5 muJy) 1.4 GHz mosaic was obtained 
at Westerbork for a 1~sq.~degr. region covering the FLSv (\cite{Mor04}). \\
The availability of both deep radio and far-infrared data is of particular 
interest, since we can exploit the well-known tight correlation 
between far-IR and radio luminosities of star-forming galaxies
(see e.g. \cite{Hel85,Gar02})
to efficiently separate radio sources triggered by AGNs. 
Very useful is also the availability of data at two radio frequencies -- 0.61 
and 1.4 GHz -- which allows us to derive the source spectral index ($\alpha$).  
This is important since different accreting regimes may display different 
spectral signatures in the radio domain. 

\section{Derivation of source spectral index}
\begin{figure}
\begin{center}
\includegraphics[width=.5\textwidth]{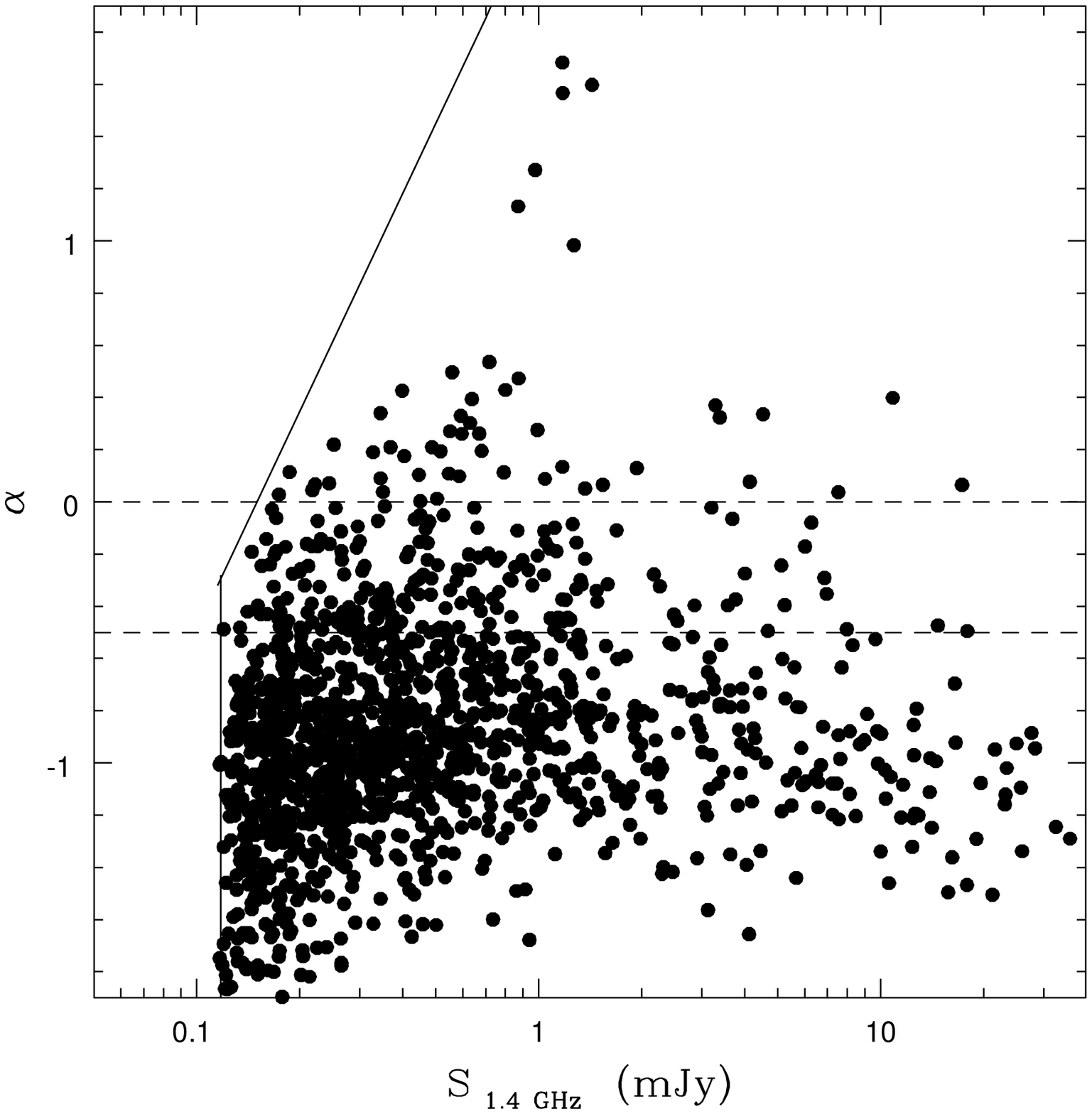}\includegraphics[width=.5\textwidth]{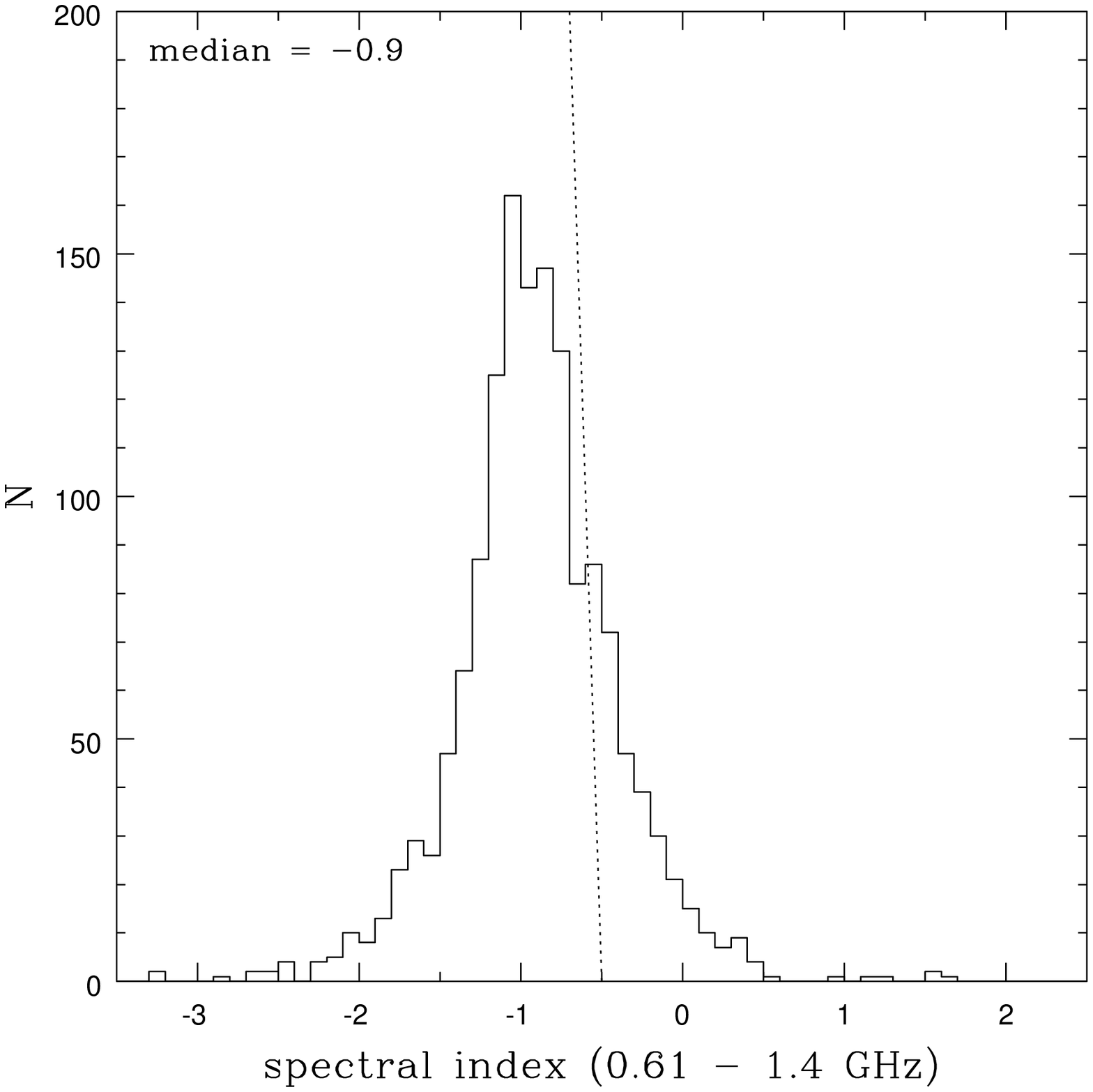}
\caption{{\it Left:} $0.61-1.4$ GHz spectral index against 1.4 GHz flux density
for the FLS sources. Horizontal lines indicate $\alpha=-0.5$ and 
$\alpha=0$, i.e. the spectral index values conventionally dividing 
steep-- from flat--spectrum sources and flat-- from inverted--spectrum sources. 
The vertical line indicates the limit of the VLA 
catalogue ($S_{\rm 1.4 GHz}\sim 0.12$ mJy), while the diagonal line 
indicate the maximum spectral index values  
to which the present work is sensitive due to the GMRT flux 
density limit ($S_{\rm 0.61 GHz}\sim 0.15$ mJy). 
{\it Right:} Distribution of FLS sources as a function of $\alpha$. 
The dotted vertical line corresponds to $\alpha=-0.5$.
\label{fig:spindex}}
\end{center}
\end{figure}

Radio sources from VLA and GMRT catalogues were cross-identified, and 
radio spectral indices between 0.61 and 1.4 GHz were derived (see 
Fig.~\ref{fig:spindex}). 
As expected, most such objects are steep-spectrum radio sources 
($\alpha<-0.5$). Nevertheless the spread in the spectral distribution is wide, 
with a significant number of flat ($-0.5<\alpha<0$) or inverted sources 
($\alpha>0$). This indicates the presence of an heterogeneous population, 
consisting in a mixture of flat/steep-spectrum AGNs and steep star-forming 
galaxies. A number of ultra-steep spectrum sources 
($\alpha<-1.1$) could be ascribed to a population of 
high-redshift galaxies.

\begin{figure}
\begin{center}
\includegraphics[width=.7\textwidth]{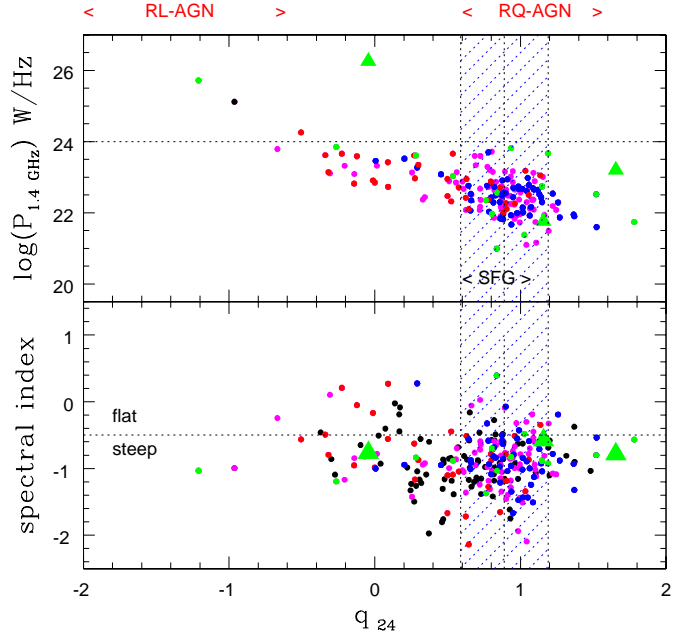}

\vspace{-1.5cm}
\caption{1.4 GHz luminosity (top) and $0.61-1.4$ GHz spectral index (bottom) 
as 
a  function of $q_{24}=log(S_{24\mu m}/S_{1.4GHz})$ for the optically identified 
FLS radio sources. 
The  expected location or star-forming galaxies is shown by the blue 
shaded region. Symbols
refer to optical spectral classification: star-forming galaxies (blue); 
early-type galaxies (red); broad/narrow emission-line AGN spectra (green 
triangles/dots); spectra showing narrow emission lines, but whose 
spectral features 
do not allow an unambiguous SFG vs. AGN classification (magenta); no spectral 
information (black). 
\label{fig:q24plot}}
\end{center}
\end{figure}

\section{Multi-wavelength analysis of FLS radio sources: radio--loud AGNs} 
The two-frequency FLS radio catalogue was cross-correlated with the 
Spitzer IRAC multi-color (3.6, 4.5, 5.8, 8.0 $\mu$m) catalogue (\cite{Lac05}), 
with the MIPS 24 $\mu$m catalogue (\cite{Fad06}) and with the 
optical spectroscopy catalogues of \cite{Mar07} and \cite{Pap06}. 
Identifying the optical counterpart of the 
sources is crucial to get information on both the galaxy redshift and 
classification (broad/narrow-line AGN, star-forming or early-type galaxy). 
An optical identification was 
found for $\sim 20\%$ of the sources, most of which have a measured redshift. 
In the following we will focus on the 
sample of optically identified sources, and use it to extract from the 
FLS a robust sub-sample of radio-emitting AGNs.\\
A first analysis of the multi-wavelength properties of the optically 
identified 
radio sources in the FLS is illustrated in Figure~\ref{fig:q24plot}, where 
we plot the 1.4 GHz radio power (top) and the 0.61-1.4 GHz spectral index 
(bottom) as a function of the so-called $q_{24}$ parameter, defined as 
the ratio between the 24 $\mu$m and the 1.4 GHz source flux density 
($q_{24}=log(S_{\rm 24mu m}/S_{\rm 1.4GHz}$). The $q_{24}$ value range allowed 
for 
star-forming galaxies is shown by the blue shaded region (\cite{Mar07}). 
Radio-loud AGNs are located at the left side of such region (corresponding to 
an excess of radio emission with respect to infrared). 
As found in 'brighter' sub-mJy samples (see e.g. the ATESP sample, 
\cite{Mig08}) we have a significant fraction of radio-loud AGNs in the FLS, 
and among them we have several flat-/inverted--spectrum sources, whose 
properties are currently under further analysis. Such radio--loud 
AGNs have mostly low radio
powers ($P_{\rm 1.4 GHz}<10^{24}$ W/Hz) and are preferentially identified with 
early-type galaxies (shown in red) or weak narrow line systems (shown in 
magenta). Such properties are consistent with those of low power (FRI-type) 
radio galaxies, typically characterized by very low accretion rates.

\begin{figure}
\begin{center}
\includegraphics[width=.7\textwidth]{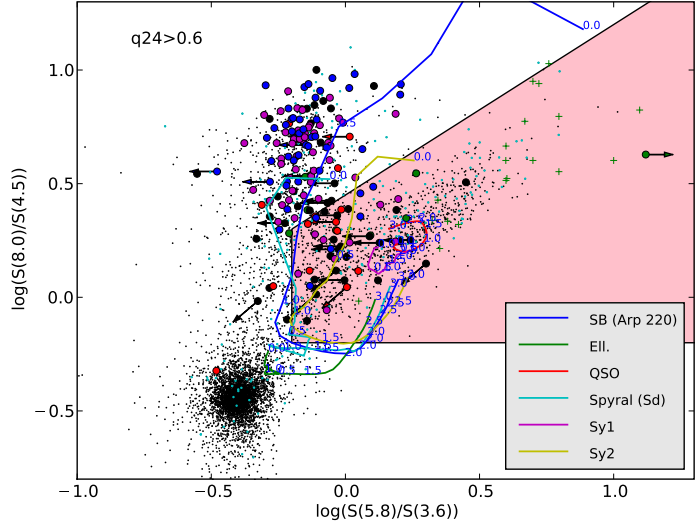}
\caption[]{IRAC color-color plot of the FLS radio sources with $q_{24}>0.6$ 
(filled symbols), i.e. for sources with infrared-to-radio ratios consistent 
with the ones of star-forming galaxies and/or radio-quiet AGNs.
Colors refer to optical spectral classification as in Figure~\ref{fig:q24plot}.
Arrows indicate upper/lower limits. 
The expected IRAC colors as a function of redshift for different source types 
are shown by different lines (see legenda in the plot). The expected
location for AGNs is highlighted in pink.
For reference we also show IRAC colors of: {\it a)} all FLS IRAC-identified 
radio sources (no optical identification selection applied, cyan dots); 
{\it b)} the entire FLS IR-selected star/galaxy population 
(no radio selection applied, black dots); and {\it c)} a sample of  
high redshift obscured (type-2) quasars 
(see Martinez-Sansigre et al. 2006, green crosses).
\label{fig:IRACplot}}
\end{center}
\end{figure}

\section{Multi-wavelength analysis of FLS radio sources: radio--quiet AGNs} 
As shown in Figure~\ref{fig:q24plot} the bulk of the FLS radio sources are 
characterized by $q_{24}>0.6$, i.e. by $q_{24}$ values
consistent with the sources being star-forming galaxies.  
Nevertheless only a fraction of such objects is optically classified as 
star-forming galaxy (blue points). A few sources are instead optically 
classified as AGNs (green points). Such AGNs represent
a first very clear direct evidence of a radio-quiet AGN population 
showing up at the radio flux levels of the FLS. 
In addition we have a few sources classified as 
early-type galaxies (red points) and a significant fraction of sources 
displaying narrow emission lines, which do not have a secure classification 
(magenta points). Among such sources we may have several other hidden 
radio--quiet AGNs.\\
In order to better disentangle radio sources triggered by star 
formation from those triggered by AGNs, we exploit the available 
IRAC colors. Figure~\ref{fig:IRACplot} shows the IRAC color-color plot for
FLS sources with $q_{24}>0.6$. As expected sources 
optically classified as star-forming galaxies (blue points) are confirmed as 
such by their IRAC colors, together with many of the sources with no secure 
optical classification (magenta points). Nevertheless we have a fraction of 
sources with typical AGN IRAC colors (objects falling in the pink region), 
which can be considered as genuine radio-quiet AGNs.  
Their radio properties ($P_{\rm 1.4 GHz}< 10^{24}$ W/Hz; steep radio spectrum, 
see Figure~\ref{fig:q24plot}) 
are fully consistent with those expected for the radio-quiet AGN population 
and, as expected, they are mostly associated to galaxies showing  emission 
lines in their optical spectra (\cite{Jar04,kuk98}). 

\section{Conclusions}
We are currently exploiting the deep radio/optical/IR information available 
for the FLS to investigate the physical properties of faint radio-selected 
AGNs. Such a study is to be included in a more general analysis of several 
radio deep fields with multi-wavelength information available, with the aim
of studying the AGN component of the sub-mJy radio fields.\\
Preliminary results show that the availability of IR colors 
proves to be very useful in disentangling star-forming galaxies from 
AGNs, allowing us to uncover a population of radio-quiet AGNs, which does not 
emerge from the radio/optical analysis only.   
The FLS radio-quiet AGN sample represents a first very clear direct evidence 
of a radio-quiet AGN population showing up at 1.4 GHz flux densities of the 
order of $S\sim 100$ $\mu$Jy, supporting the idea that radio-quiet AGNs are not 
necessarily radio silent. 
This result is very promising, possibly indicating that the bulk of the 
radio-quiet AGN population may be discovered 
going to deeper radio flux densities than the FLS flux 
limit.

\end{document}